\documentclass{pasj00}
\draft
\SetRunningHead{Kodaira et al.}{Two LAEs beyond $z$ 6 in the SDF}
\Received{}
\Accepted{}
\Published{}
\begin{document}
\title{THE DISCOVERY OF TWO LYMAN$\alpha$ EMITTERS BEYOND REDSHIFT 6
       IN THE SUBARU DEEP FIELD\altaffilmark{*, $\dagger$}}
\author{  K. Kodaira          \altaffilmark{1},
          Y. Taniguchi        \altaffilmark{2},
          N. Kashikawa        \altaffilmark{3},
          N. Kaifu            \altaffilmark{3},
          H. Ando             \altaffilmark{3},
          H. Karoji           \altaffilmark{4}, 
          M. Ajiki            \altaffilmark{2}, \\
          M. Akiyama          \altaffilmark{4},
          K. Aoki             \altaffilmark{4},
          M. Doi              \altaffilmark{5},
          S. S. Fujita        \altaffilmark{2},
          H. Furusawa         \altaffilmark{4}, 
          T. Hayashino        \altaffilmark{6}, \\
          M. Imanishi         \altaffilmark{3}, 
          F. Iwamuro          \altaffilmark{7},
          M. Iye              \altaffilmark{3},
          K. S. Kawabata      \altaffilmark{3},
          N. Kobayashi        \altaffilmark{4}, \\
          T. Kodama           \altaffilmark{3},
          Y. Komiyama         \altaffilmark{4},
          G. Kosugi           \altaffilmark{4},
          Y. Matsuda          \altaffilmark{6},
          S. Miyazaki         \altaffilmark{4}, \\
          Y. Mizumoto         \altaffilmark{3}, 
          K. Motohara         \altaffilmark{5}, 
          T. Murayama         \altaffilmark{2},
          T. Nagao            \altaffilmark{2},
          K. Nariai           \altaffilmark{8}, \\
          K. Ohta             \altaffilmark{7},
          Y. Ohyama           \altaffilmark{4},
          S. Okamura          \altaffilmark{9, 10}, 
          M. Ouchi            \altaffilmark{9},
          T. Sasaki           \altaffilmark{4}, \\
          K. Sekiguchi        \altaffilmark{4},
          K. Shimasaku        \altaffilmark{k},
          Y. Shioya           \altaffilmark{2}, 
          T. Takata           \altaffilmark{4}, \\
          H. Tamura           \altaffilmark{6}, 
          H. Terada           \altaffilmark{4}, 
          M. Umemura          \altaffilmark{11},
          T. Usuda            \altaffilmark{4}, \\
          M. Yagi             \altaffilmark{3}, 
          T. Yamada           \altaffilmark{3}, 
          N. Yasuda           \altaffilmark{3}, \\ and
          M. Yoshida          \altaffilmark{12} 
}

\altaffiltext{1}{The Graduate University for Advanced Studies (SOKENDAI),\\
        Shonan Village, Hayama, Kanagawa 240-0193}
\altaffiltext{2}{Astronomical Institute, Graduate School of Science,
          Tohoku University, \\ Aramaki, Aoba, Sendai 980-8578}
\altaffiltext{3}{National Astronomical Observatory of Japan,
          2-21-1 Osawa, Mitaka, Tokyo 181-8588}
\altaffiltext{4}{Subaru Telescope, National Astronomical Observatory,\\
          650 N. A'ohoku Place, Hilo, HI 96720, USA}
\altaffiltext{5}{Institute of Astronomy, Graduate School of Science,
          University of Tokyo, \\ 2-21-1 Osawa, Mitaka, Tokyo 181-0015}
\altaffiltext{6}{Research Center for Neutrino Science, Graduate School of 
  Science, Tohoku University, \\ Aramaki, Aoba, Sendai 980-8578}
\altaffiltext{7}{Department of Astronomy,  Graduate School of Science,
        Kyoto University, \\ Kitashirakawa, Sakyo, Kyoto 606-8502}
\altaffiltext{8}{Department of Physics, Meisei University, 2-1-1 Hodokubo,
        Hino, Tokyo 191-8506}
\altaffiltext{9}{Department of Astronomy, Graduate School of Science,
        University of Tokyo, Tokyo 113-0033}
\altaffiltext{10}{Research Center for the Early Universe, Graduate School
        of Science, \\ University of Tokyo, Tokyo 113-0033}
\altaffiltext{11}{Center for Computational Physics, University of Tsukuba,\\
        1-1-1 Tennodai, Tsukuba 305-8571}
\altaffiltext{12}{Okayama Astrophysical Observatory, National Astronomical
  Observatory of Japan, \\ Kamogata-cho, Asakuchi-gun, Okayama 719-0232}


\maketitle

\footnotetext[*]{Based on data collected at the Subaru Telescope, which is 
                 operated by the National Astronomical Observatory of Japan.}
\footnotetext[$\dagger$]{This work has been done with a collaboration between
       the Subaru Deep Field Project led by the association of   
       builders of the Subaru telescope and the common-use, 
       Intensive Program (S02A-IP-2) led by Y. Taniguchi.}

\begin{abstract}
We have performed a deep optical imaging survey using a narrowband
filter ($NB921$) centered at $\lambda =$ 9196 \AA ~ together with $i^\prime$
and $z^\prime$  broadband filters covering an 814 arcmin$^2$ area of the
Subaru Deep Field.
We obtained a sample of 73 strong  $NB921$-excess objects 
based on the following two color criteria;
$z^\prime - NB921 > 1$ and $i^\prime - z^\prime > 1.3$.
We then obtained optical spectroscopy of nine objects 
in our $NB921$-excess sample, and identified at least two  
Ly$\alpha$ emitters at$z=6.541 \pm 0.002$ and $z=6.578 \pm 0.002$, each of which
shows the characteristic sharp cutoff 
together with the continuum depression at wavelengths
shortward of the line peak.
The latter object is more distant than HCM-6A at $z=6.56$
and thus this is the most distant known object found so far.
These new data allow us to estimate the first meaningful lower limit of
the star formation rate density beyond redshift 6;
$\rho_{\rm SFR} \sim 5.2 \times 10^{-4} M_\odot$ yr$^{-1}$ Mpc$^{-3}$.
Since it is expected that the actual density is higher by a factor of
several than this value, our new observation reveals 
that a moderately high level of star formation activity
already occurred at $z \sim$ 6.6.
\end{abstract}

\section{INTRODUCTION}

Probing the star formation activity in galactic or subgalactic systems
in early universe is important for understanding both the history of galaxies
and the origin of cosmic reionization (e.g., Loeb, Barkana 2001).
Recent advance in deep optical imaging capability with 8-10 m class telescopes
enabled new searches for star-forming galaxies beyond redshift 5.
In particular, imaging surveys using narrow-passband filters have
proved to be an efficient way to find such galaxies (Ajiki et al. 2002;
Cowie, Hu 1998, Hu et al. 1996, 1999, 2002; Kudritzki et al. 2000;
Rhoads et al. 2001; Steidel et al. 2000; Taniguchi et al. 2003).
Indeed, the most
distant Ly$\alpha$ emitter (LAE) known to date, HCM 6A at $z=6.56$ was 
discovered by using this technique (Hu et al. 2002).
However, surveys for emission-line galaxies with narrow-band filters
have an intrinsic limitation in redshift coverage and hence the
survey volumes are not often large enough to ensure robust 
for success. In order to increase survey volumes
and to reach faint limiting magnitude,
we need wide-field CCD cameras on 8-10 m class telescopes. 
Suprime-Cam (Miyazaki et al. 2002) mounted at the prime focus of
the 8.2 m Subaru Telescope (Kaifu 1998) on Mauna Kea, Hawaii, provides a unique
opportunity for wide-field (a $34^\prime \times 27^\prime$ field of view),
narrowband imaging surveys for emission-line objects at high redshift.

The most distant LAE, HCM 6A at $z = 6.56$ (Hu et al. 2002), is 
gravitationally amplified by a factor of 4.5 by the foreground cluster 
of galaxies Abell 370 at $z =0.37$.
Although the help of any gravitational lensing is highly useful
in investigating faint high-$z$ objects (Ellis et al. 2001; Hu et al. 2002),
it is also important to search
for high-$z$ LAEs in a so-called blank field for an unbiased study.
In an attempt to find star-forming objects at $z \approx 6.6$ 
in such a blank field,
we have carried out a very deep optical imaging survey in the 
Subaru Deep Field (SDF)
centered at $\alpha$(J2000) = $13^{\rm h} ~ 24^{\rm m} ~ 21.4^{\rm s}$ and
$\delta$(J2000) = $+27^\circ ~ 29' ~ 23''$ (e.g., Maihara et al. 2001;
Ouchi et al. 2003; Kashikawa et al. 2003).
In this Letter, we report on our discovery of two Ly$\alpha$ emitters
at $z \approx$ 6.5 -- 6.6.

\section{OBSERVATIONS}

\subsection{Optical Imaging}

In this survey, we used the narrow-passband filter, NB921, centered on
$\lambda_{\rm c}$ = 9196 \AA ~ with a passband of $\Delta\lambda$(FWHM)
= 132 \AA, corresponding to a redshift range between 6.508 and 6.617
for the Ly$\alpha$ emission.
Optical imaging was made in the $i'$, $z'$, and $NB921$ bands
on a central $34'\times 27'$ area of the SDF with Suprime-Cam. 
Our direct imaging was obtained during several observing runs
between 2001 April and 2002 May. The total integrations time was
4.7 hr, 5.8 hr, and 5.0 hr for $i^\prime$, $z^\prime$, and
$NB921$, respectively. The data reduction procedures are the same
as those given in Yagi et al. (2002).
The PSF FWHM of the final images is $0.''90$.
Source detection and photometry are performed using
SExtractor version 2.1.6 (Bertin, Arnouts 1996).
The limiting magnitude (AB) for a 5$\sigma$ detection
on a $1.''8$ diameter aperture is 26.9, 26.1, and 25.7 for
$i^\prime$, $z^\prime$, and $NB921$, respectively.

For each object detected in the $NB921$ image,
$i^\prime$, $z^\prime$, and $NB921$ magnitudes are measured on a common
aperture of $1.''8$ diameter.
In total, 50,449 objects are detected down to $NB921=25.7$ 
(the $5\sigma$ limiting magnitude). The effective area used to search 
for $NB921$-excess objects is 814.3 arcmin$^2$.
The FWHM half--power points of the filter correspond to a co-moving depth along
the line of sight of 40.9 $h_{0.7}^{-1}$ Mpc.
Thus a total volume of 202,000 $h_{0.7}^{-3}$ Mpc$^{3}$ is probed 
in our $NB921$ image.

We select candidates of $z\simeq 6.6$ LAEs imposing two criteria,
$z'-NB921 > 1$ and $i'-z' > 1.3$, on the above objects.
There are 404 objects satisfying the first criterion alone.
The latter criterion is used to reduce contamination from
foreground objects. Objects at $z \approx 6.6$ have sharp breaks because
of the strong Ly$\alpha$ absorption at this redshift 
(Songaila, Cowie 2002) and are expected to exceed the
adopted i-z criterion while low redshift galaxies invariably do not.
This latter criterion is applied only to objects brighter than
$i'=28.0$ ($\simeq 2\sigma$ limiting magnitude),
and we retain all the objects (satisfying $z'-NB921 > 1$)
with $i' \ge 28.0$ in order not to miss possible faint LAEs.
These selection procedures yield eventually a photometric
sample of 73 LAE candidates.
Note that among the objects with $i' \ge 28$, those brighter
than $z'=26.7$ ($3\sigma$ limiting magnitude) satisfy
automatically $i'-z'>1.3$, but we cannot obtain a meaningful
constraint on $i'-z'$ for those with $z' \ge 26.7$.

\subsection{Optical Spectroscopy}

In order to investigate the nature of LAE candidates found in our
optical imaging survey, we obtained optical spectroscopy of nine
objects in our LAE candidate sample using
the Subaru Faint Object Camera And Spectrograph (FOCAS: Kashikawa
et al. 2002) on 2002 June 7-9 under a
0.$^{\prime\prime}$55 - 1.$^{\prime\prime}$10 seeing condition.
This spectroscopic sample contains
the brightest three objects with $NB921 < 24$. However, the remaining
six objects are randomly selected from the photometric sample and 
they have $NB921 \approx 25$ -- 25.5. 

Our optical spectroscopy was made with the 300 lines mm$^{-1}$
grating and an O58 order cut filter. The wavelength coverage was
between 6000 \AA ~ to 10000 \AA ~
with a pixel resolution of 1.34 \AA.
The use of an 0.8 arcsec-wide slit gave a spectroscopic resolution of
9.0 \AA ~ at 9200 \AA ~ (or $R \simeq 1020$).
The spatial resolution was 0.$^{\prime\prime}$3 pixel$^{-1}$
by 3-pixel, on-chip binning.
Spectroscopy of the brightest object in our LAE sample ($NB921$ = 23.01)
was obtained in the long-slit spectroscopy mode.
The exposure time was 1800 seconds. This source was quickly identified
as an [O {\sc iii}]$\lambda$5007 emitter at $z \approx 0.84$.
Other eight objects were observed in the multi-object spectroscopy (MOS) mode.
We chose two fields that contained as many LAE candidates as possible;
hereafter field 1 and field 2.
We succeeded in obtaining spectra of
three targets in Field 1 and five in Field 2.
The same grating, filter, and slit as those in the long-slit mode
were used in this MOS mode.
We obtained twelve and six 1800s exposures
for Field 1 and Field 2, respectively. We also obtained the spectrum of
a standard star Hz 44 for flux calibration.

\section{RESULTS}

Our optical spectroscopy (see Figure 1) indicates that 
at least two objects are well-defined LAEs between $z = 6.5$ and $z = 6.6$
because their emission-line shapes show the sharp cutoff on the UV side
together with the continuum depression at wavelengths
shortward of the line peak.
Two more objects also show the sharp cutoff at wavelengths
shortward of the line peak. However, their continuum magnitudes
are so faint that we cannot see firm evidence for the continuum break.
Although they are probable candidates of LAEs at $z \approx 6.5$ -- 6.6,
we do not include them in the later discussion\footnote{Spectroscopic 
properties of
other five objects are as follows. Three objects show a single 
emission line around $\lambda$9200 \AA ~ in our optical spectra.
Since they show an almost symmetrical 
emission-line profile, we cannot identify
them as LAEs solely basing upon our optical spectroscopy; they may be either
an [O {\sc ii}]$\lambda$3727 source at $z \approx 1.46$ or a
LAE at $z \approx 6.6$ (e.g., Stern et al. 2000).
It seems also worthwhile noting that a symmetric emission-line
profile does not necessarily rule out the case of Ly$\alpha$
emission (e.g., ESO 350-IG038 in Kunth et al. 1998).
The remaining two sources are confirmed to be 
[O {\sc iii}]$\lambda$5007 emitters at $z \approx 0.84$.}.
In the spectra of the well-defined two LAEs, SDF J132415.7+273058
in Field 2 and SDF J132418.3+271455 in Field 1
[panels (a) and (b), respectively], 
continuum emission appears to be present at wavelengths 
longer than the Ly$\alpha$ peak.
Combining these two spectra and applying a 3-pixel smoothing,
we obtain the average spectrum in panel (c). 
The average continuum flux density between 1200 \AA ~ and 1210 \AA ~ is
$(1.7 \pm 4.1) \times 10^{-19}$ ergs s$^{-1}$ cm$^{-2}$ \AA$^{-1}$
while that between 1219 \AA ~ and 1226 \AA ~ is
$(6.4 \pm 3.7) \times 10^{-19}$ ergs s$^{-1}$ cm$^{-2}$ \AA$^{-1}$.
This difference is regarded as evidence for the continuum break in the 
spectra of these two LAEs. 

The redshifts of the two LAEs are estimated from the peak of 
the Ly$\alpha$ emission line and the results are given in Table 1
together with line widths.
It is noted that SDF J132418.3+271455 ($z = 6.578$)
is more distant than HCM 6A at $z=6.56$ (Hu et al. 2002) because error of
our redshift measurement is $\pm$ 0.002.
Therefore, SDF J132418.3+271455 is the most distant LAE known to date.
Thumbnail images of the two objects are shown in Figure 2. 
The $NB921$ images
reveal that only SDF J132415.7+273058 is spatially extended;
its angular diameter (FWHM) is estimated as 1.2 arcsec.
Correcting for the seeing spread (0.9 arcsec), we obtain an angular
diameter of 0.8 arcsec, corresponding to 4.4 $h_{0.7}^{-1}$ kpc
at $z=6.541$. 

\section{DISCUSSION}

The SDF is a so-called blank field
and there is no apparent cluster of galaxies  known to date
at low and intermediate redshifts in our field.
Since, further, the lensing effect is expected to be small in this field,
our survey results allow us to perform a simple statistical analysis
of star formation activity in the investigated volume
even though our sample is not large.

First, we estimate the star formation rate of the LAEs at $z \approx 6.6$
by using
the relation $SFR({\rm Ly}\alpha)  = 9.1 \times 10^{-43} L({\rm Ly}\alpha) ~
M_\odot {\rm yr}^{-1}$ (Kennicutt 1998; Brocklehurst 1971).
The observed Ly$\alpha$ flux, the Ly$\alpha$ luminosity,
and the star formation rate, SFR(Ly$\alpha$), of each LAE are summarized 
in Table 1  where a flat universe with $\Omega_{\rm matter} = 0.3$,
$\Omega_{\Lambda} = 0.7$, and $h=0.7$ with $h = H_0/($100 km s$^{-1}$
Mpc$^{-1}$) is adopted.
The average SFR obtained for the two LAEs is 7.1 $\pm$ 2.0
$h_{0.7}^{-2} ~ M_\odot$ yr$^{-1}$, being comparable to those of LAEs at 
$z \simeq$ 5.1 -- 5.8 (e.g., Ajiki et al. 2002).
It should be mentioned that the SFRs estimated above are lower limits
because it is quite likely that a blue half or more of the Ly$\alpha$ emission
may be absorbed by H {\sc i} gas and dust grains in the galaxy itself and by
the intergalactic H {\sc i} gas (Miralda-Escud\'e 1998; 
Miralda-Escud\'e, Rees 1998; Cen, McDonald 2002).
The SFR based on the Ly$\alpha$ luminosity tends to be underestimated
by a few times or more than that based on the UV luminosity (see also
Hu et al. 2002).
Indeed, using the average UV continuum flux density between 
1219 \AA ~ and 1226 \AA ~
for the combined spectrum shown in panel (c) of Figure 1 together with
the relation (Kennicutt 1998),
$SFR({\rm UV}) = 1.4 \times 10^{-28}L_{\nu} ~~ M_{\odot} ~ {\rm yr}^{-1}$,
where $L_\nu$ is in units of ergs s$^{-1}$ Hz$^{-1}$,
we obtain an average value of 
$SFR({\rm UV})\simeq 22  h_{0.7}^{-2} ~ M_\odot$ yr$^{-1}$ for the two LAEs.
As for SDF J132415.7+273058, we obtained $J$-band imaging using 
the InfraRed Camera and Spectrograph (IRCS: Kobayashi et al. 2000)
on the Subaru Telescope 
on 2002 July 15. The integration time was 6480 seconds (the detail
will be given elsewhere). The $J$ magnitude (AB) is estimated as $\simeq$ 24.9. 
This photometry allows us to estimate the star formation rate at
$\lambda_{\rm rest}$ = 1650 \AA ~ for SDF J132415.7+273058; $SFR$(UV) $\simeq
36 h_{0.7}^{-2} ~ M_\odot$ yr$^{-1}$, which is even higher by a factor of 
four than $SFR$(Ly$\alpha$). This is suggestive of dust obscuration
at the bluest wavelengths.

Now we can estimate the total star formation rate of 73 LAEs in our 
photometric sample using the equivalent width of NB921 flux.
We have identified one LAE (SDF J132415.7+273058) in the three brightest
candidates and the other (SDF J132418.3+271455) in the faint sample.
Because of small number statistics, it seems modest to assume that
approximately 22\% (=2/9) of
73 LAE candidates are real LAEs at $z \approx$ 6.5 - 6.6
in any magnitude range; note that
95 \% confidence level based on the random
sampling hypothesis ranges 8 \% - 49 \%. 
If we assume that all the 73 LAE candidates are true LAEs 
at $z \approx 6.5$ -- 6.6,
we obtain nominally a total star formation rate of 
$SFR_{\rm total}^{\rm nominal} = 475 h_{0.7}^{-2} ~ M_\odot$ yr$^{-1}$.
Provided that approximately 22\% of 
73 LAE candidates are real LAEs at $z \approx$ 6.5 - 6.6,
we can estimate the total star formation rate,
$SFR_{\rm total} \simeq 0.22 \times SFR_{\rm total}^{\rm nominal} 
\simeq 105  h_{0.7}^{-2} ~ M_\odot$ yr$^{-1}$.
Given the survey volume, 202,000 $h_{0.7}^{-3}$ Mpc$^{3}$, we thus
obtain a star formation rate density of $\rho_{\rm SFR} \simeq 5.2 \times 10^{-4}
h_{0.7} ~ M_\odot$ yr$^{-1}$  Mpc$^{-3}$.
It should be reminded here again that
we apply neither any reddening correction nor integration by assuming
a certain luminosity function for LAEs. Further, we note that there are two more
probable LAE candidates in our spectroscopic sample.
Therefore, our estimate should be regarded as a robust and first meaningful
lower limit for the star formation rate density beyond $z=6$.
Let us compare this value with previous estimates in Figure 3;
note that we convert all the previous estimates to those in the cosmology
adopted in this Letter.
In conclusion, the present study shows unambiguously that the moderate
star formation activity already occurred in the early universe beyond $z=6$.


\bigskip

We would like to thank  the Subaru Telescope staff for
their invaluable assistance. We would also like to thank
Prof. Hy Spinrad both for useful comments for the earlier version 
of this paper and for encouragement.
We also thank the referee, Daniel Stern, and another anonymous referee
for their useful comments and suggestions.



\newpage


\begin{table}
\caption{Properties of the two LAEs \label{tbl-1}}
{\scriptsize
\begin{center}
\begin{tabular}{cccccccccccc}
\hline\hline
No. &
Name\footnotemark[a] &
\multicolumn{3}{c}{Optical AB Magnitude} &
$z$ &
\multicolumn{2}{c}{FWHM\footnotemark[b]} &
EW\footnotemark[c] &
$F$(Ly$\alpha$) &
$L$(Ly$\alpha$) &
$SFR$(Ly$\alpha$) \\
 &
 &
$i^\prime$ &
$z^\prime$ &
$NB921$ &
 &
\AA &
km/s &
\AA &
10$^{-17}$ ergs/s/cm$^2$ &
$10^{42}$ ergs/s &
$M_\odot$ yr$^{-1}$ \\
\hline
1 & SDF J132415.7+273058 & 27.48 & 25.82 & 23.99 & 6.541 & 10.9 & 357
  & $160^{+520}_{-70}$  & $2.06 \pm 0.18$
  & $10.02 \pm 0.09$ & $9.1 \pm 0.8$ \\
2 & SDF J132418.3+271455 & 28.52 & 26.66 & 24.98 & 6.578 & $<9.0$ & $<290$
  & $330^{+\infty}_{-200}$  & $1.13 \pm 0.05$
  & $5.58 \pm 0.25$ & $5.1 \pm 0.2$ \\
\hline
\end{tabular}
\end{center}

$^a$ The sky position, $\alpha$(J2000) and $\delta$(J2000),
     is given in the name.\\
$^b$ Full width at half maximum of the Ly$\alpha$ emission line.\\
$^c$ Equivalent width of the Ly$\alpha$ emission line at observed frame.\\
}
\end{table}


\begin{figure}
\begin{center}
\FigureFile(80mm,50mm){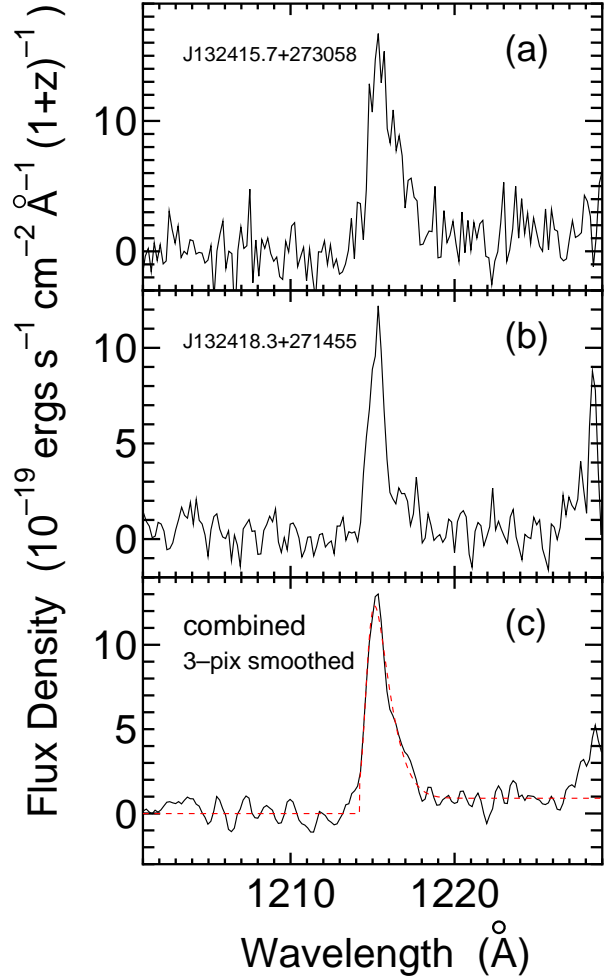}
\end{center}
\caption{
The rest-frame UV spectra of two LAE
candidates between 1200 \AA ~ and 1230 \AA.
The spectra of SDF J132415.7+273058 and SDF J132418.3+271455 are shown in 
panel (a) and (b), respectively. The combined spectrum of these two LAEs
is shown in panel (c). A trial of the profile fitting with a combination
between emission and absorption is shown by red line.}
\label{fig:fig1}
\end{figure}

\begin{figure}
\begin{center}
\FigureFile(80mm,50mm){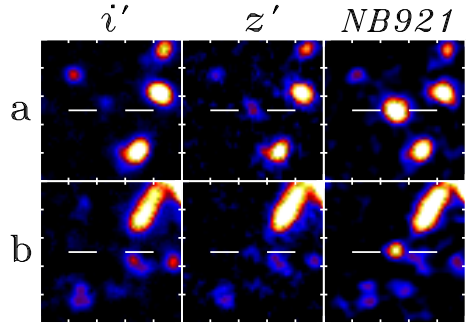}
\end{center}
\caption{
Thumbnail images of the two well-defined LAEs, (a) SDF J132415.7+273058 
and (b) SDF J132418.3+271455.
The size of each image is $10^{\prime\prime}\times 10^{\prime\prime}$,
and north is up and east is left.}
\label{fig:fig2}
\end{figure}

\begin{figure}
\begin{center}
\FigureFile(80mm,50mm){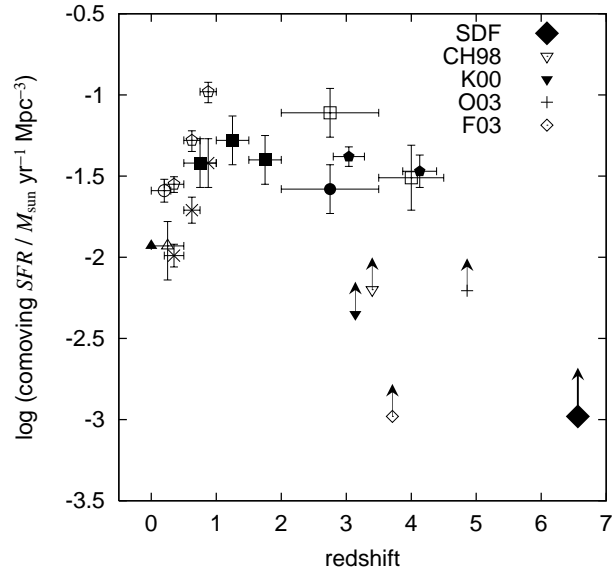}
\end{center}
\caption{
The star formation rate density ($\rho_{\rm SFR}$) as a function of 
redshift $z$. Our estimate at $z \approx 6.6$  (large filled diamond)
is shown together with  the results of previous Ly$\alpha$ searches at 
$z \sim 3$ - 5 (CH98 = Cowie \& Hu 1998, K00 = Kudritzki et al. 2000, 
F03 = Fujita et al. 2003, and O03 = Ouchi et al. 2003).
The previous investigations are shown by 
filled triangle (Gallego et al. 1996), open triangle (Treyer et al. 1998),
open circle (Tresse \& Maddox 1998), stars (Lilly et al. 1996), 
open pentagons (Hammer et al. 1999), 
filled squares (Connolly et al. 1997), filled circles (Madau 1998),
and open squares (Pettini et al. 1999).
Other results for Lyman break galaxies between $z$ = 3 - 4 
are also shown by filled pentagons (Steidel et al. 1999).}
\label{fig:fig3}
\end{figure}

\end{document}